\documentclass{aa}

\usepackage{graphicx}
\usepackage{txfonts}
\usepackage{xcolor}
\usepackage{amsmath}
\usepackage{todonotes}
\usepackage{threeparttable}
\usepackage{booktabs}
\usepackage[labelfont=bf, labelsep=period]{caption}
\usepackage{physics}
\usepackage{amssymb}
\usepackage{array}
\usepackage{float}
\usepackage{amsfonts}
\usepackage{macro_ref}

\usepackage{hyperref}

\begin{document}

   \title{A scalable and accurate framework for self-calibrating null depth retrieval using neural posterior estimation}
   \titlerunning{First results of NPE in nulling interferometry}

   \author{Baoyi Zeng\inst{1}
          \and
          Marc-Antoine Martinod\inst{2}
          \and
          Denis Defr{\`e}re\inst{2}
          \fnmsep
          }

   \institute{Faculty of Engineering Science, KU Leuven, Kasteelpark Arenberg 1, 3001 Leuven, Belgium
         \and
             Institute of Astronomy, KU Leuven, Celestijnenlaan 200D, 3001 Leuven, Belgium
             }

   \date{Received XX; accepted XX}

\abstract{Accurate null depth retrieval is critical in nulling interferometry. However, achieving accurate null depth calibration is challenging due to various noise sources, instrumental imperfections, and the complexity of real observational environments. These challenges necessitate advanced calibration techniques that can efficiently handle such uncertainties while maintaining a high accuracy.}
{This paper aims to incorporate machine-learning techniques with a Bayesian inference to improve the accuracy and efficiency of null depth retrieval in nulling interferometry. Specifically, it explores the use of neural posterior estimation (NPE) to develop models that overcome the computational limitations of conventional methods, such as numerical self-calibration (NSC), providing a more robust solution for accurate null depth calibration.}
{An NPE-based model was developed, with a simulator that incorporates real data to better represent specific conditions. The model was tested on both synthetic and observational data from the LBTI nuller for evaluation.}
{The NPE model successfully demonstrated improved efficiency, achieving results comparable to current methods in use. It achieved a null depth retrieval accuracy down to a few $10^{-4}$ on real observational data, matching the performance of conventional approaches while offering significant computational advantages, reducing the data retrieval time to one-quarter of the time required by self-calibration methods.}
{The NPE model presents a practical and scalable solution for null depth calibration in nulling interferometry, offering substantial improvements in efficiency over existing methods with a better precision and application to other interferometric techniques.}

\maketitle

\section{Introduction}

The study of exoplanets is of great importance, as it sheds light on the formation and evolution of planetary systems \citep{turrini2015role}. However, imaging exoplanets presents several challenges because of the great distance and the extreme contrast ratio between the brightness of the planet and its host star, which typically ranges from $10^{-4}$ to $10^{-10}$, depending on the wavelength \citep{schworer2015predicting}. This issue is compounded by the fact that stars and their exoplanets are often separated by only a few milli-arcseconds when observed from Earth. This proximity necessitates sophisticated techniques to differentiate between the two sources of light. 

Offering a promising solution to these challenges, the technique of nulling interferometry allows for the direct detection of exoplanets, by canceling out the starlight through destructive interference and benefiting from the higher angular resolution offered by long-baseline interferometry, to bring orbiting exoplanets into view \citep{bracewell1978detecting}. 
Two requirements for nulling interferometry are the maintenance of a deep null and the accurate calibration of the null depth. 
This is challenging because, during observation, the measured null depth faces noise sources including atmospheric disturbance and instrumental imperfection. 

The classical method for null depth calibration uses a calibrator star with known properties and geometry \citep{colavita2009keck}. By observing the calibrator star under the same conditions assumed to remain constant, instrumental errors can be corrected. However, this method’s accuracy depends on the calibrator star’s known characteristics, with the assumption of unchanged observation conditions often being unreliable \citep{norris2022optimal}. Numerical self-calibration (NSC, \citealt{hanot2011improving}), on the other hand, mitigates the need for a calibrator star. It uses statistical analysis of the observational data distribution to extract the necessary scientific information, proving to be more precise than classical methods, significantly reducing the impact of systematic errors \citep{mennesson2011new}.
It has enabled several scientific achievements on spectroscopic binaries \citep{serabyn2019nulling}, exozodiacal dust \citep{defrere2016nulling, defrere2021hosts, ertel2020}, and stellar diameters \citep{norris2020first, lagadec2021glintsouth, martinod2021scalable}. The NSC approach relies on a model of the instrument (that includes the parameter of the self-calibrated null depth) and Monte Carlo simulations to reproduce the statistical distributions of the nulled signal. Classical algorithms of model fitting such as the maximum likelihood estimation are used to find the best self-calibrated parameters of the model. Despite its advantages, this method faces two practical limitations. First, common NSC implementations fit marginal (histogram-based) distributions of the nulled signal and auxiliaries, which discards cross-channel and temporal correlations that carry information about astrophysical signal and systematics. Second, its scalability and computational efficiency are limited by the large amount of simulations and model fitting required to process the various targets and pointings obtained in a given sequence of observations. The accuracy of self-calibration strongly depends on the complexity of the instrumental models: the more realistic and detailed the models, the more extensive the simulations are at every iteration of the fitting algorithm. This makes the process computationally expensive and time-consuming, particularly for large observing programs.

With advancements in machine learning, innovative approaches to parameter inference have emerged, offering new tools for null depth calibration. Among these is neural posterior estimation (NPE, \citealt{papamakarios2016fast, lueckmann2017flexible}), which combines Bayesian inference with artificial neural networks. Bayesian inference has been widely used for parameter retrieval in exoplanet studies, especially for atmospheric retrieval. 
The most common methods of Bayesian inference in this context include Markov chain Monte Carlo (MCMC, \citealt{metropolis1953equation, hastings1970monte}) and nested sampling methods \citep{skilling2004nested, skilling2006nested}, which help to estimate the values of interest and the uncertainties of model parameters. 
However, there are significant challenges associated with MCMC and nested sampling techniques. A primary limitation is their heavy computational burden: depending on model complexity, this process can demand hundreds to thousands of hours of computationally parallelizable time \citep{himes2022accurate}. This computational cost also makes it impractical to validate statistical reliability through repeated runs \citep{vasist2023neural}. 
Moreover, these methods typically demand an explicit and tractable likelihood function, a limitation highlighted, for example, in the context of simulation-based inference by \cite{cranmer2020frontier}. This restricts simulation models to relatively simple or deterministic forms and limits their flexibility. While the burden tends to increase as models incorporate more nuisance parameters, the central difficulty lies not in the number of parameters per se, but in the inefficiency, validation challenges, and likelihood dependence of sampling-based inference.

The similarity in data reduction challenges faced during both null depth calibration and atmospheric retrieval suggests that Bayesian inference could be a viable method for null depth retrieval. Naturally, it would encounter issues similar to those seen in atmospheric retrieval. With the help of machine learning, NPE offers solutions to these challenges \citep{fluke2020surveying}. It works by generating synthetic data through simulations, which are then used to train a neural network to approximate the posterior distribution of null depth. By leveraging deep learning, NPE can handle complex data and generate fast, accurate inferences about the null depth. One of the key advantages of NPE is its ability to bypass the need for an explicit likelihood function \citep{nixon2020assessment, cranmer2020frontier}. This likelihood-free inference makes NPE particularly well-suited for complex systems such as nulling interferometers. Another key advantage lies in the amortized inference process, which enhances efficiency by performing computationally intensive tasks upfront \citep{vasist2023neural}. Once trained, these models can quickly infer new data points without repeating the entire simulation process. 

Although not yet widely adopted, several studies have demonstrated that NPE is a tool with significant potential. \cite{vasist2023neural} present a pivotal study demonstrating NPE's ability to significantly reduce inference time while maintaining both accuracy and reliability. By utilizing a complex radiative transfer model, the study shows that NPE can generate accurate posterior approximations within seconds, positioning it as a promising method for scalable atmospheric retrievals. Moreover, the application of NPE in parameter estimation across other fields, including gravitational wave science, has further highlighted its potential \citep{dax2023neural, chua2020learning}.

This study aims to explore the application of NPE on nulling data, seeking to improve the accuracy and efficiency of null depth calibration. The structure of this paper is as follows. Section \ref{sec: methods} outlines the core concepts and workflow of NPE, along with the model setup used in this study. In Section \ref{sec: results} and \ref{sec: discussion}, we discuss our findings before analyzing NPE’s limitations and making proposals for potential future directions. The thesis concludes with summaries presented in Section \ref{sec: conclusions}.

\section{Design of the normalizing flow}
\label{sec: methods}

\subsection{Simulation-based inference}

The NPE method is part of the simulation-based inference (SBI) framework, which has emerged as a robust tool for performing statistical inference when the likelihood function is intractable, therefore also referred to as likelihood-free inference \citep{cranmer2020frontier, brehmer2022simulation}. Unlike traditional sampling-based techniques such as MCMC, which require the explicit calculation of the likelihood, SBI circumvents this requirement by using simulations to approximate the posterior distribution. 
Serving as the core of SBI, a simulator can be defined as a computer program that takes a parameter vector \( \theta \) as input, samples latent variables or internal states \( z \) based on transition probabilities \( p_i(z_i \vert \theta, z_{<i}) \), where the index \( i \) denotes the position in the sequence of latent variables, before eventually generating a data vector \( x \sim p(x \vert \theta, z) \) as output.
The produced data $x$ corresponds to the simulated observation. The objective of inference is to deduce the parameters of interest $\theta$, based on the observed data $x$. 

SBI offers significant advantages over traditional Bayesian methods, particularly in its capacity for likelihood-free inference. By using simulators to generate synthetic data, SBI effectively approximates posterior distributions without needing an analytically intractable likelihood function \citep{Tejero-Cantero2020sbi}. This approach is particularly advantageous in dealing with more complex and realistic physics models \citep{dax2021real}. SBI also employs summary statistics and distance measures to directly compare simulated and observed data. Moreover, the amortized inference process trains surrogate models to quickly infer new data points, providing substantial computational savings, especially in high-dimensional settings \citep{schmitt2023detecting}.

With the development of deep learning, integrating neural networks into SBI has provided even more robust solutions, including NPE (\citealt{blum2010non, papamakarios2016fast, lueckmann2017flexible}). It involves training a neural network on simulated data and corresponding parameters. The network learns to model the conditional distribution of the parameters given the data, effectively learning a surrogate posterior distribution. This method allows for the direct handling of data with complex dependencies that traditional methods struggle with.
Therefore, it provides the scalability and accuracy needed for self-calibration in nulling interferometry.

\subsection{Workflow of neural posterior estimation}

\begin{figure}[H]
    \centering
    \includegraphics[width=0.95\linewidth]{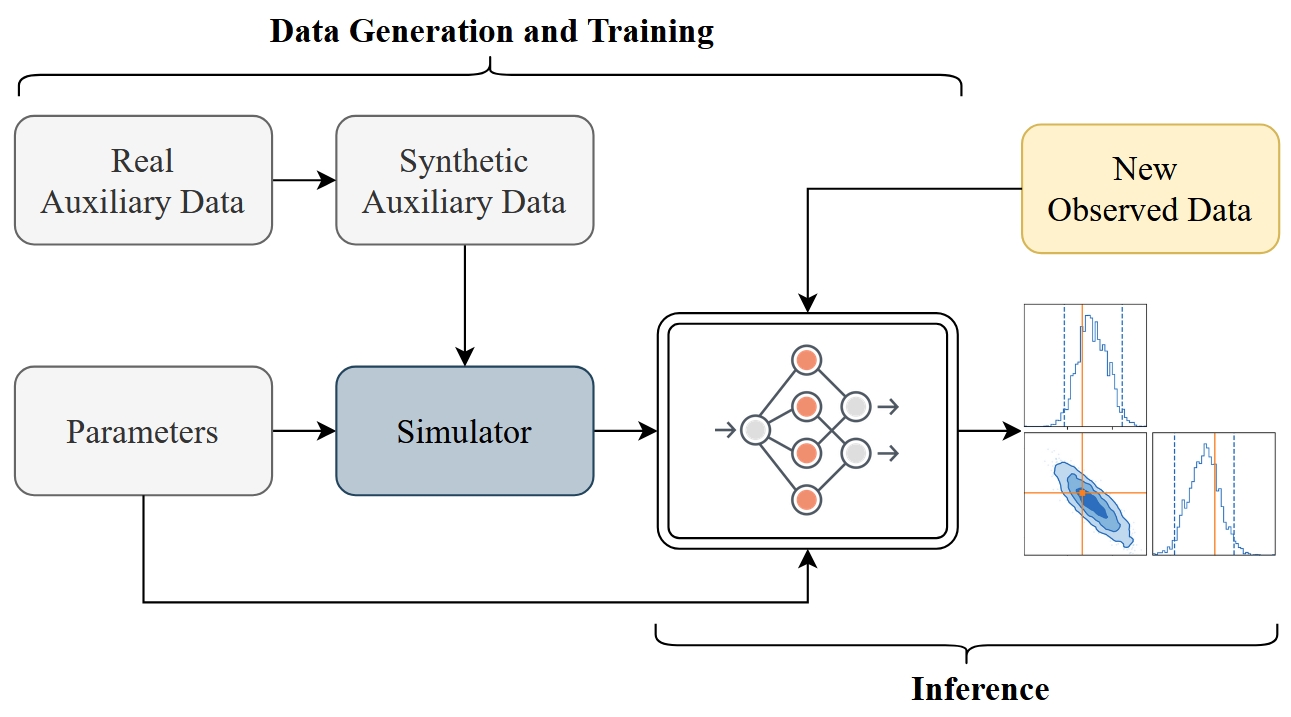}
    \caption{Workflow diagram NPE. The process consists of data generation, neural network training, and inference-making. }
    \label{fig:workflow_npe}
\end{figure}

The workflow of NPE involves several key steps,
which can be summarized as simulation, training, and inference. The model structure we used is illustrated in Figure \ref{fig:workflow_npe}, which visually demonstrates the entire workflow.

\paragraph{Simulation}

During simulation, a comprehensive dataset is generated, which consists of pairs of model parameters of interest \( \theta \) and corresponding simulated observations \( x \). Formally, a stochastic model is defined to generate synthetic \( x \) given \( \theta \). Since our model is designed to target the Large Binocular Telescope Interferometer (LBTI, \citealt{hinz2016overview,ertel2020overview}), we construct the simulator using a similar underlying physical model. 
The LBTI is an AO-equipped N-band Bracewell nulling interferometer combining the light from the two 8-m primary apertures of the Large Binocular Telescope (LBT). The physical model involves generating a synthetic sequence of flux measurements as a function of time $I_-(t)$, which can then be compared with the actual measured sequence \citep{defrere2016nulling, mennesson2011new}. After taking the short integration time $T$ into account, the average synthetic flux measurement over $T$ can be represented as

    \begin{equation}
    \label{eq:IminusSimulator}
    \begin{aligned}
    I_{-} \simeq 
    & \, I_1 + I_2  \\
    & + 2 \sqrt{I_1 I_2} \; \Bigg| \frac{1 - N_a}{1 + N_a} \Bigg| \; \cos\left(\frac{2 \pi}{\lambda} \Delta \text{OPD}\right) \; (1 - 0.5 \sigma_{\epsilon}^2 + 0.125 \sigma_{\epsilon}^4) \\
    & + B,
    \end{aligned}
    \end{equation}

\noindent where $I_1$ and $I_2$ are the photometric intensities of two beams of the interferometer, $B$ is the thermal background measurement, and \(\Delta OPD\) (optical path difference) causes the differential phase. These four terms are averaged over the integration time $T$. $V$ represents the source visibility at baseline, which is linked to the value of interest, source null depth, by $N_a = (1-|V|)\,/\,(1+|V|)$. $\sigma_{\epsilon}$ represents the fringe blurring term, defined as the standard deviation of the phase over time $T$. This term is included to account for the varying differential phase over the integration time. By addressing the varying terms over time, we can model the distribution of auxiliary data \(I_1\), \(I_2\), \(B\), and \(\sigma_\epsilon\) by histograms, instead of dealing with a sequence of time series. 
This marginalization is a modeling choice that does not exploit temporal or cross-channel correlations; our current NPE realization therefore shares the correlation-loss limitation of standard NSC. Correlation-preserving remedies are discussed in the section~\ref{sec:Limitations}.

We generate our data pairs \((\theta, f(\theta))\) using a simulator, following the physical process described in Equation~\ref{eq:IminusSimulator}. The parameters of interest \(\theta\) include the astrophysical null $N_a$, the mean \(\mu_{OPD}\) and standard deviation \(\sigma_{OPD}\) of the OPD. They are generated by sampling from the prior distributions, as shown in Table~\ref{tab:theta prior}.
We select these priors based on previous experience with both the instrument and the observed star.
These priors are set to fit the current dataset, specifically the observational data from $\beta$~Leo. 
However, they can be adjusted to accommodate stars with different ranges of values, if other datasets require it.
The simulator can also produce synthetic auxiliary data by sampling from their respective priors. However, in the current model, we directly inject observed auxiliary data distributions into the model. 
These terms are sourced from the empirical cumulative distribution function of measured values to closely follow the statistical properties of real observations. 
This approach partially simulates the real data generation process. It operates under the sole assumption that photometries remain constant during an observing block (OB).

\vspace{0.3 cm}

    \begin{table}[H]
    \centering
    \normalsize
    \caption{Prior distributions for parameters of interest (OPD in~nanometer).}
    \begin{tabular}{c c}
    \hline
    Parameter & Prior \\
    \hline
    $N_a$ & $\mathcal{U}(0, 0.01)$ \\
    $\mu_{OPD}$ & $\mathcal{N}(300, 150)$ \\
    $\sigma_{OPD}$ & $\mathcal{U}(100, 300)$ \\
    \hline
    \end{tabular}
    \label{tab:theta prior}
    \end{table}

\paragraph{Training}

Once the dataset is prepared, the next step is to train a neural network to approximate the posterior distribution \( p(\theta\vert x) \). The NPE method uses normalizing flows to achieve this. Normalizing flows \citep{papamakarios2021normalizing, kobyzev2020normalizing} are a series of invertible transformations $T$ applied to a simple base distribution, such as a standard normal distribution, to obtain a complex target distribution, while making it easy to sample and evaluate. 

After defining the normalizing flow, a flow-based model \( p_\phi(\theta \vert x) \) is trained on the generated data, which in the case of NPE is a neural network where \(\phi\) represents its parameters. This model approximates the posterior \( p(\theta\vert x) \). To train the neural network, we employ variational inference, which approximates complex posterior distributions \citep{blei2017variational}. 
The used metric is the Kullback-Leibler (KL) divergence \citep{jiang2018approximate}. The optimization process minimizes the KL divergence between the true posterior \( p(\theta\vert x) \) and the approximated posterior \( p_\phi(\theta\vert x) \), effectively allowing the model to learn a distribution that best represents the posterior. This is achieved by maximizing the Evidence Lower Bound (ELBO), which provides a tractable objective for training the model. Gradient-based optimization methods, including stochastic gradient descent, are then employed to maximize the ELBO, thus refining the parameters \( \phi \) to yield an accurate posterior estimate. 

\paragraph{Inference}

After training, the neural network can be used to perform inference on new observations \( x_{\text{obs}} \). The process of inference involves a forward pass through the trained normalizing flows to obtain samples from the posterior distribution \( p_{\phi}(\theta\vert x_{\text{obs}}) \). This step is computationally efficient and can be executed in a matter of seconds. Therefore, NPE is considered scalable, as it is both fast and reusable. The efficiency gained from amortization means that once the neural network is trained, it can be reused for multiple sets of data without the need for retraining, significantly reducing computational costs.

\subsection{Model setup}

In our study, the posterior estimator \( p_{\phi}(\theta|x) \) is designed using a neural autoregressive flow \citep{huang2018neural}. This estimator is constructed with three sequential transformations, each parametrized by a multilayer perceptron, which consists of eight hidden layers: three with 128 units and five with 64 units. The activation function used is ELU \citep{clevert2015fast}. During the training process, the neural network is trained for 32 epochs, each consisting of 128 batches with a batch size of 64. The training process aims to minimize the expected negative log density of the posterior over the training set. The learning rate is set to \(10^{-3}\) and the weight decay is set to \(10^{-2}\), using the AdamW optimizer for parameter updates \citep{loshchilov2017decoupled}. For code implementation, we used the LAMPE package\footnote{https://github.com/probabilists/lampe} which was built upon the PyTorch framework and specifically designed for simulation-based inference. We also made use of the GRIP package\footnote{https://github.com/mamartinod/grip} which was a toolbox designed for the implementation of NSC \citep{martinod2024generic}. 

The hyperparameters of the neural network were optimized via a grid search, evaluating combinations of batch size, epochs per batch, and overall number of epochs. Specifically, the grid search parameters include batch sizes ranging in [32, 256], number of epochs ranging in [32, 64], and number of batches per epoch ranging in [64, 256]. 
Additionally, various neural network architectures were tested, with the number of hidden layers ranging in [5, 12] and their dimensions in [32, 256]. The optimal configuration was selected based on achieving the lowest validation loss while maintaining stable training performance.

\section{Retrieval and validation results}
\label{sec: results}
We aim to investigate the efficiency of our NPE implementation to retrieve null depth measurements.
We start by investigating its capability of parameter retrieval on mock data, then on real astronomical data.
Finally, we compare the NPE and NSC through their results on the real data.

\subsection{Retrieval on synthetic data}
\label{subsec: synthetic retrieval}

We first test the ability of our NPE implementation to retrieve the null depth on mock data.
To do so, we used a set of fixed parameters to generate this dataset which has been given to the NPE implementation, which has to find the former.
The parameters to be retrieved $\theta$ are as follows: $N_a$ = 0.0063, $\mu_{OPD}$ = 300~nm, $\sigma_{OPD}$ = 250~nm.
We selected these values to align with typical observed data from the LBTI nuller, and auxiliary data were sourced from an observing block (OB) of LBTI's observations of the star $\beta$~Leo, ensuring that test conditions closely mimic real-world scenarii. 
We also applied the NSC method from the GRIP package \citep{martinod2024generic} to the same set of mock data to allow a direct evaluation of the two approaches under identical conditions.

{
\begin{table}[H]
\centering
\normalsize
    \caption{Results of null depth retrieval on synthetic data ($\mu_{OPD}$ and $\sigma_{OPD}$ in nanometer).}
    \label{tab:benchmark}
    \renewcommand{\arraystretch}{1.5}
    \begin{tabular}{c c c c}
        \hline
        Parameter & Truth & NPE retrieval & NSC retrieval\\
        \hline
        $N_a$ & 0.0063 & $0.0061^{+0.0002}_{-0.0003}$ & $0.0065^{+0.0008}_{-0.0009}$\\
        $\mu_{OPD}$ & 300 & $308.82^{+11.50}_{-11.89}$ & $296.56^{+27.96}_{-31.01}$\\
        $\sigma_{OPD}$ & 250 & $241.75^{+8.16}_{-7.58}$ & $240.46^{+21.12}_{-18.10}$\\
        \hline
    \end{tabular}
    \begin{tablenotes}
        \footnotesize
        \item Notes: Medians of posterior distributions are used as point estimates. 
        Superscripts and subscripts give the $16^\mathrm{th}$ and $84^\mathrm{th}$ percentiles (i.e., the 68\% credible interval).
    \end{tablenotes}
\end{table}
}

    \begin{figure}[H]
    \centering
        \hspace{0.15 cm}\includegraphics[width=0.95\linewidth]{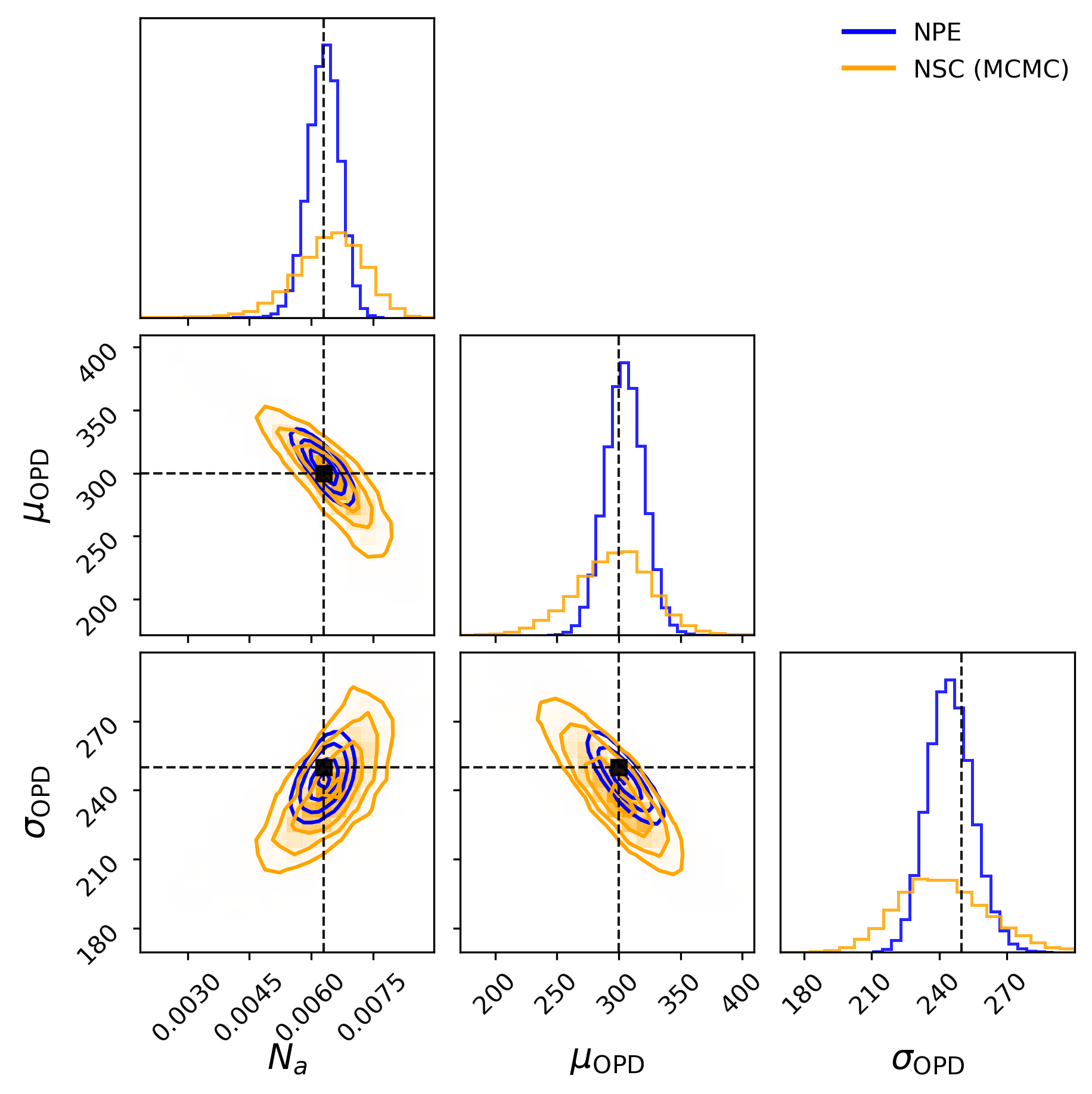}
    \caption{Comparison of the posterior distributions obtained with the NPE (blue) and NSC (orange) methods on the same synthetic dataset.
    The retrieved parameters are \( N_a \), \( \mu_{OPD} \), and \( \sigma_{OPD} \).
    The histograms along the diagonal show the marginalized distributions of each parameter, while the off-diagonal plots display the joint distributions between pairs of parameters.
    The black dashed crosses mark the ground truth values.
    }
    \label{fig:benchmark}
    \end{figure}

The outcomes of the inference are presented in Figure~\ref{fig:benchmark} and summarized in Table~\ref{tab:benchmark}. 
The NPE model demonstrates precise parameter estimation, as indicated by its tight marginalized distributions.
Such an outcome occurs as long as the prior distributions used to train the neural network are consistent with the priors used to create the mock data: the parameters of the data can be learned from the prior distributions, and the latter do not simulate too many unrealistic situations during the training process as well. The joint distributions reveal clear and well-defined contours, suggesting that the model effectively captures the underlying parameter dependencies.
Figure~\ref{fig:benchmark} directly compares the posterior distributions obtained using NPE (blue) and NSC (orange).
Both methods recover values that are consistent with the ground truth, but the NPE posteriors are visibly narrower and more concentrated around the true parameters.
This strengthens the case that our NPE implementation not only recovers the true parameters as reliably as NSC but can also provide sharper constraints.

\begin{figure}[H]
    \centering
    \includegraphics[width=0.7\linewidth]{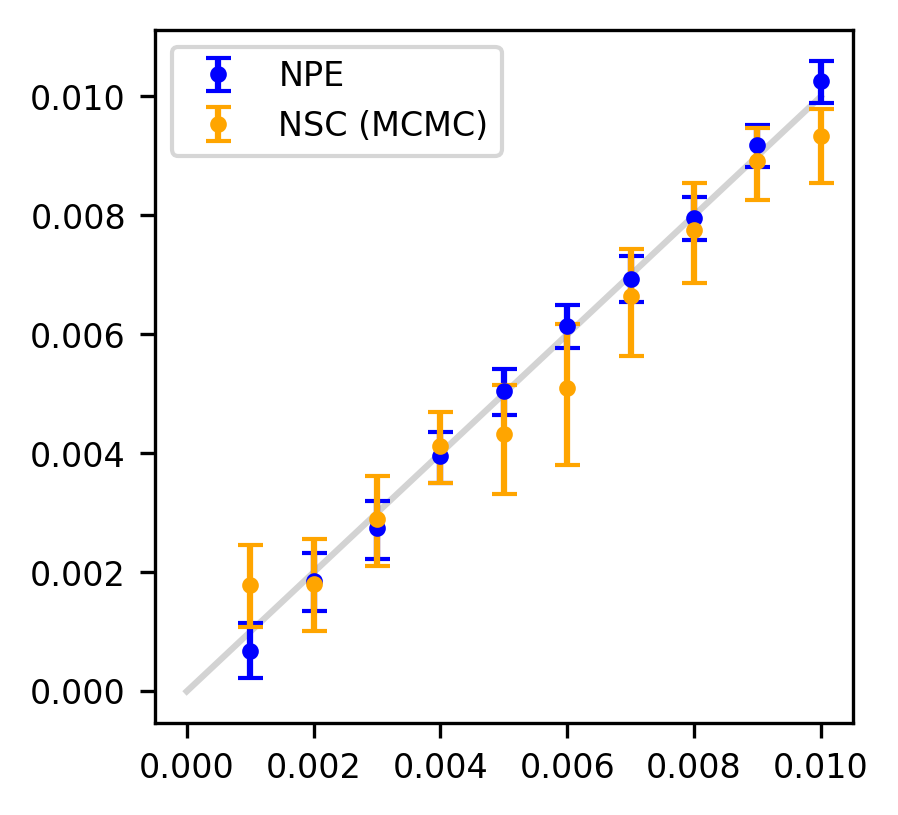}
    \caption{
    Retrieval of the null depth using NPE (blue) and NSC via MCMC (orange) for mock datasets with true null depths ranging from 0.001 to 0.010 and fixed $\mu_{OPD} = 300$~nm and $\sigma_{OPD} = 250$~nm. 
    The gray line indicates the one-to-one correspondence. 
    Error bars correspond to the $16^{\mathrm{th}}$ and $84^{\mathrm{th}}$ percentiles of the posterior distributions.
    }
    \label{fig:null_depth_scan}
\end{figure}

To further evaluate the robustness of both methods, we extended the test to a range of synthetic datasets with null depths ranging from 0.001 to 0.010, while keeping $\mu_{OPD}$ and $\sigma_{OPD}$ fixed.
In the case of the NPE, we use the same neural network as the one which retrieved the parameters described in Figure~\ref{fig:benchmark}.
Figure~\ref{fig:null_depth_scan} shows the retrieved null depths for both NPE and NSC in this range. The mean absolute errors of the retrievals by the NPE and NSC methods are $1.5 \times 10^{-4}$ and $4.1 \times 10^{-4}$, respectively.
The NPE results closely follow the true values and show consistently narrower error bars, whereas the NSC retrievals exhibit further deviations of the medians from the truth and larger uncertainties.
This confirms that the NPE approach provides tighter and more stable parameter constraints across different regimes of null depth.

\subsection{Retrieval on LBTI data}

In order to evaluate NPE's ability to retrieve null depth accurately, we compare the results of the NPE model with the current methods used by the LBTI nuller, which is a method combining the NSC technique and the use of a reference star \citep{defrere2016nulling}. 
This approach achieves an accuracy down to a few \(10^{-4}\). 
A comparison of the accuracy as well as the computation time is conducted, allowing us to assess NPE's capability to match or even exceed the accuracy of existing techniques while potentially offering improvements in computational efficiency.

The data we used to test the performance of NPE was obtained from the LBTI nuller, specifically from commissioning observations made in February 2015 \citep{defrere2016nulling}. 
Since the LBTI uses a reference star to measure and calibrate the instrumental error, a typical observational plan includes interleaving pointings of the science and reference stars. 
The data we used included 49 OBs, divided into six pointings, consisting of observations on the science target $\beta$~Leo and three calibrator stars (Table~\ref{tab:obs_log}).
One OB consists of one main dataset with the nulled signal, and three auxiliary datasets: one with the thermal background and one with each beam open individually.
The main dataset also contains the root mean square values of the residuals of the fringe tracker within an integration time, used to quantify $\sigma_\epsilon$ in Equation~\ref{eq:IminusSimulator}, which is considered as auxiliary data.

\begin{table}[H]
    \centering
    \caption{Observation log of the night of February 8, 2025. A total of 49 OBs were acquired.}
    \begin{tabular}{ccc}
        \hline
         Star & CAL/SCI    & Number of OBs  \\
         \hline
         HD 104979      & CAL   & 8\\
         $\beta$ Leo    & SCI   & 8\\
         HD 109742      & CAL   & 9\\
         $\beta$ Leo    & SCI   & 8\\
         HD 108381      & CAL   & 8\\
         HD 109742      & CAL   & 8\\
         \hline
    \end{tabular}
    \label{tab:obs_log}
\end{table}

In the data reduction process for LBTI data, we use the measured values of the auxiliary data from the first OB of each pointing and inject them into the simulator.
The simulator is used to train the neural network, which is then used to infer the astrophysical null depth for all the OBs of the current pointing.
We adopt this method because the statistical properties of measurements may vary significantly across different pointings, as pointings on the science object are interleaved with pointings on calibrator stars. This approach ensures that the observations can be accurately modeled by the generated data.

    \begin{figure}[H]
    \centering
            \includegraphics[width=\linewidth]{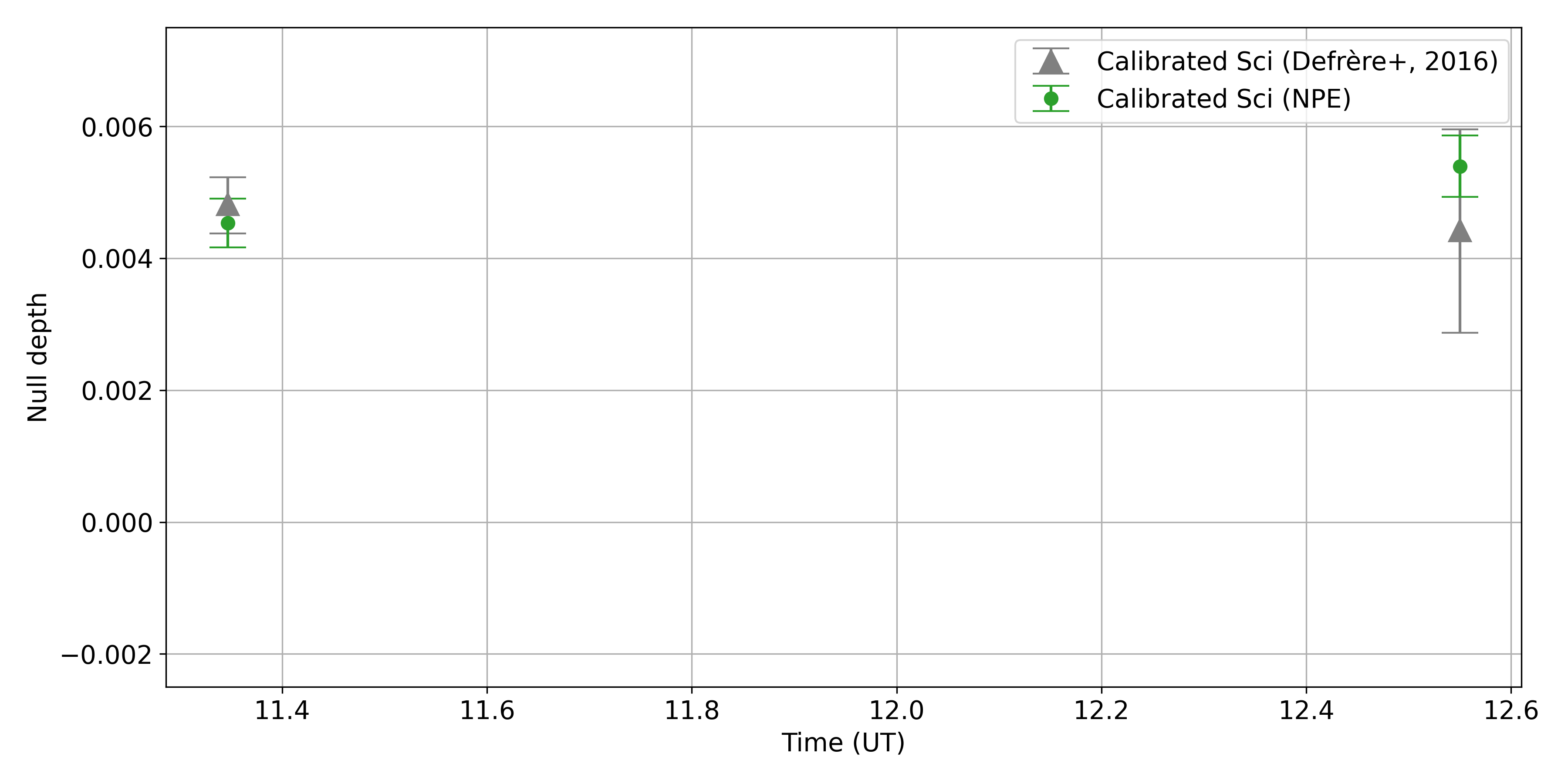}

    \caption{Null depth of the science star $\beta$~Leo over time retrieved by the NPE model, calibrated using reference stars. }
    \label{fig:astro}
    \end{figure}

    \begin{table}[H]
    \centering
    \caption{Null depth of $\beta$~Leo retrieved by the NPE model, compared with the results from \cite{defrere2016nulling}.}
    \begin{tabular}{ccc}
        \hline
         Source of estimates & First pointing  & Second pointing   \\
         \hline
          \cite{defrere2016nulling}  & 0.478\% $\pm$ 0.050\%  & 0.439\% $\pm$ 0.156\%\\
         NPE model & 0.453\% $\pm$ 0.037\%  & 0.539\% $\pm$ 0.046\%\\
         \hline
    \end{tabular}
    \label{tab:astro}
\end{table}

After performing the NPE, we used the results of the calibrator stars to compute systematic errors or the so-called null floor to get better calibration accuracy, following the same methodology as in \cite{defrere2016nulling}.
Since the null floor remains relatively stable during observations, we calculated a transfer function (TF) using a weighted average of all three calibrator measurements. 
The error calculation for the TF also followed the approach in \cite{defrere2016nulling}. While this method is now considered outdated \citep{ertel2018}, it was adopted here to ensure consistency and enable a meaningful comparison.
The calibrated null depth of the science star is obtained by subtracting the null floor from the null depth retrieved by the model. 
Figure~\ref{fig:astro} provides a visualization of the final results of null depth calibration, comparing the outcomes obtained by NPE with those from classical methods. Specific numerical values are summarized in Table~\ref{tab:astro}.
While both methods yield consistent results, the NPE approach outperforms the classical method by a factor of 3 in the second pointing. 
Additionally, the NPE approach suggests potential variability in the astrophysical null between the two pointings, although only at the ~2-sigma level. This would not be a surprise given the ~12-degree difference in parallactic angles between the two measurements.
\vspace{0.5 cm}

\subsection{Validation}

We performed three diagnostic checks to ensure the accuracy and reliability of the posterior approximations. These included the prior predictive check, the posterior predictive check \citep{sinharay2003posterior, conn2018guide}, and the coverage ratio.
All three checks were applied to the synthetic data retrieval described in Section~\ref{subsec: synthetic retrieval}, hereafter the validation data set, as the calculation of the coverage ratio requires ground truth values, which are unavailable when making inferences on observational data.
The prior predictive check evaluates the quality of the priors by comparing simulated data, generated from the prior distribution of model parameters, with the validation set (specifically, the histograms of the null depth). This ensures that the priors are reasonable. 
The posterior predictive check assesses whether the model's posterior distribution can accurately capture the underlying distribution of the inferred parameters. In this check, simulated data are generated using parameters sampled from the posterior and compared with the validation set. 

    \begin{figure}[ht]
        \begin{tabular}{ll}
            \hspace{-1em}\includegraphics[width=0.27\textwidth]{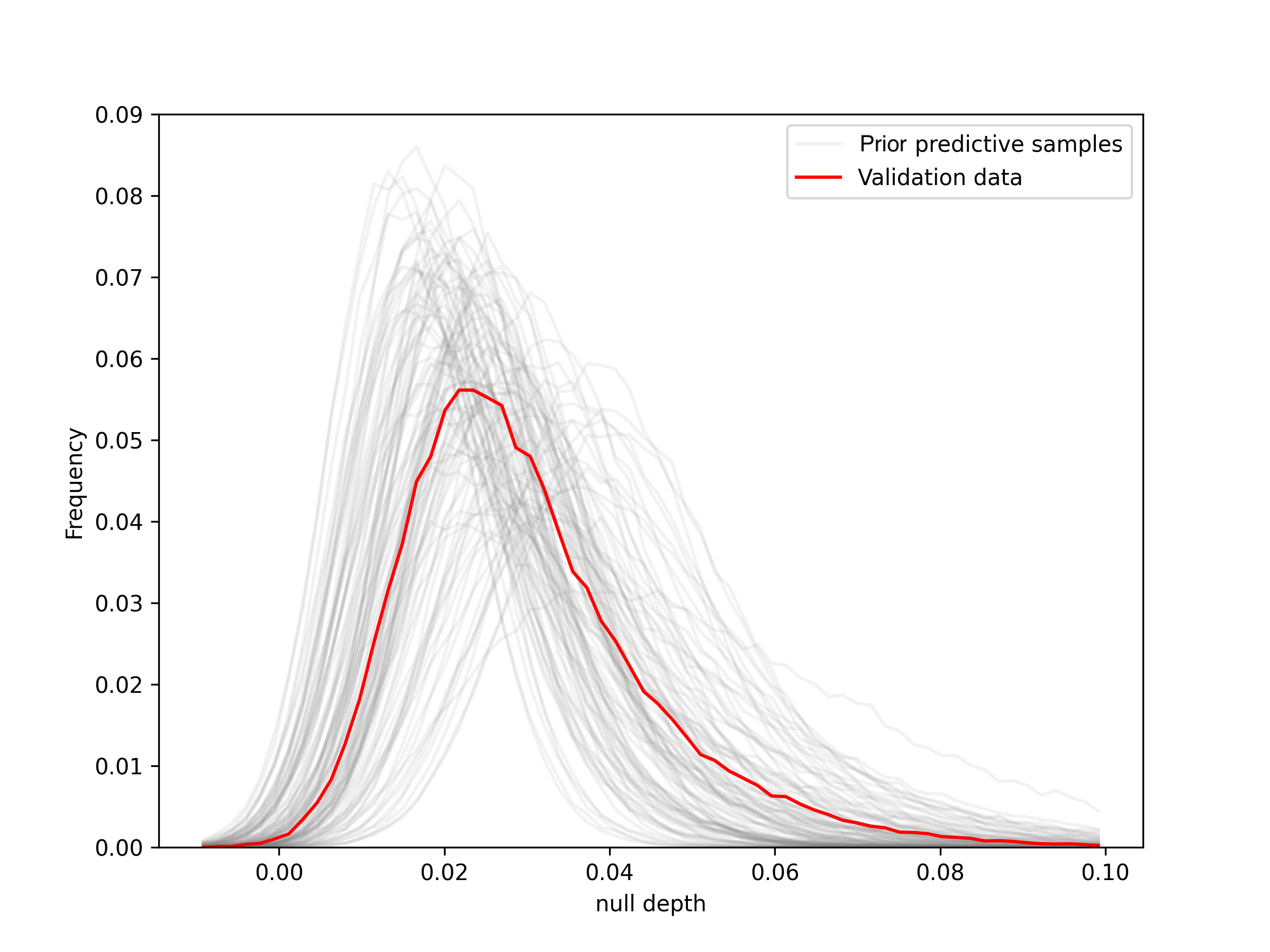}& \hspace{-2.5em}\includegraphics[width=0.27\textwidth]{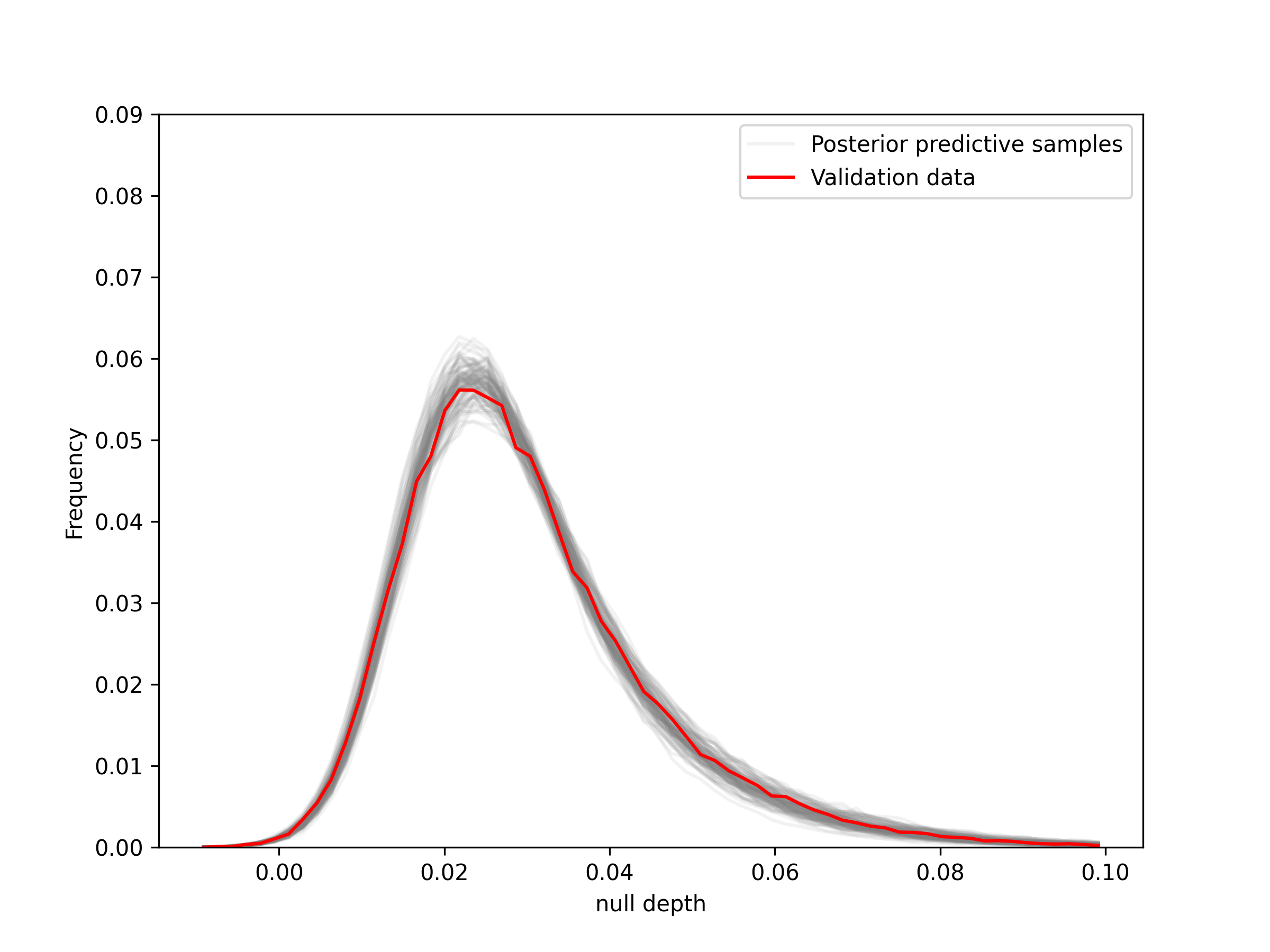} \\
        \end{tabular}
    \caption{Prior (left) and posterior (right) predictive check for the NPE model.
    Left: The grey semi-transparent lines represent data generated from the prior predictive distribution, indicating the range of data the model might encounter during training. The red solid line represents the validation data to be inferred by the model, specifically for the measured null. 
    Right: The grey semi-transparent lines represent the posterior predictive samples for each model. The red solid line represents the validation data during the inference. }
    \label{fig:ppc}
    \end{figure}

The results of prior and posterior predictive checks are presented in Figure~\ref{fig:ppc}. 
The prior predictive distribution closely approximates the real data, effectively capturing the observed values. This suggests that the prior distribution of the parameters $\theta$ is well-justified and consistent with the actual observations. 
Similarly, the posterior predictive distribution aligns closely with the validation data, indicating that the model is able to effectively capture the underlying patterns. 
By incorporating the observed auxiliary data directly, the model circumvents potential uncertainties associated with the simulation of these data points. This approach also minimizes variability encountered during training, thereby enabling the model to generate a more accurate posterior distribution. Training on data that closely resembles the actual observations enhances the model's ability to produce precise and reliable predictions.

The last performed check is the calculation of the expected coverage ratio.
It is employed to assess whether the approximation for the posterior is conservative or over-confident \citep{hermans2021trust}.
The metric evaluates the performance of posterior estimators by determining the proportion of true parameter values that lie within the credible intervals of the estimated posterior distributions. 
It involves generating a set of i.i.d. samples from the joint distribution of parameters and validation data, then estimating the posterior distributions for each sample. 
The expected coverage probability can be defined as

\begin{equation}
    \mathbb{E}_{p(\theta, x)} \left[ \mathbb{I} \left( \theta \in \Theta_{\hat{p}(\theta|x)}(1 - \alpha) \right) \right],
\end{equation}

\noindent where \(\Theta_{\hat{p}(\theta|x)}(1 - \alpha)\) represents the \((1 - \alpha)\) highest posterior density region of the estimated posterior distribution \(\hat{p}(\theta|x)\), and \(\mathbb{I}\) is the indicator function that equals 1 if the true parameter value \(\theta\) is within this credible region and 0 otherwise. 
A perfectly calibrated posterior should have an expected coverage probability that matches the nominal credibility level \(1 - \alpha\). In this case, the coverage curve overlaps with the diagonal line in the visualization.
Deviations from this ideal indicate either overconfidence (coverage less than \(1 - \alpha\)), where the coverage curve falls below the diagonal line, or conservativeness (coverage greater than \(1 - \alpha\)), where the coverage curve exceeds the diagonal line. In both cases, these deviations signal potential issues with the posterior estimations. 

    \begin{figure}[H]
    \centering
        \includegraphics[width=0.75\linewidth]{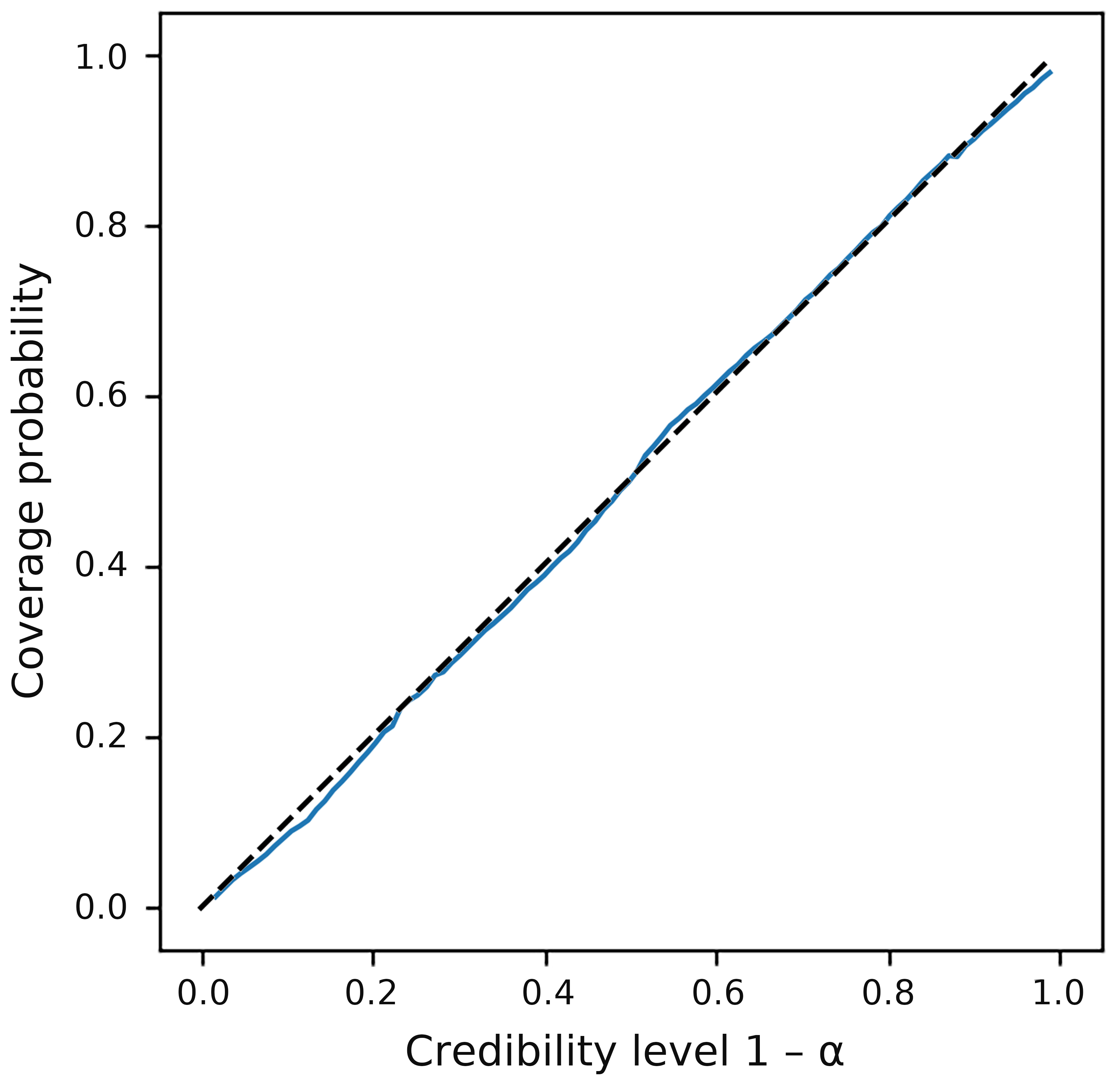}
    \caption{Coverage probability of the NPE model. The plot depicts the coverage probability as a function of the credibility level \(1 - \alpha\). 
    The solid blue lines represent the model's coverage probabilities, while the dashed line indicates the ideal scenario in which the coverage probability exactly matches the credibility level. 
    In this ideal case, the parameter values \(\theta\) used to generate the data \(x\) are contained in the \(1 - \alpha\) highest posterior density regions of the approximate posteriors \(p_\phi(\theta | x)\) exactly \(1 - \alpha\)\% of the time.}
    \label{fig:coverage}
    \end{figure}

Figure~\ref{fig:coverage} depicts the coverage ratio of the NPE model. 
The coverage curve of the model closely fits the diagonal line, indicating that the posterior distributions produced by NPE are well-calibrated, neither over-dispersed nor under-dispersed. This demonstrates that the neural network effectively serves as a surrogate posterior. 

\section{Discussion}
\label{sec: discussion}

In this study, we designed an NPE model that gives consistent results with the reference paper. The quality check also validated the reliability of the results.
In the following sections, we discuss the results we have obtained in Section \ref{sec:discuss model}, investigate the trade-off between accuracy and efficiency in Section \ref{sec:trade-off}, address the encountered limitations while exploring potential future directions for this research in Section \ref{sec:Limitations}.

\subsection{Achievement of the NPE model}
\label{sec:discuss model}

The NPE model demonstrated high accuracy when tested on synthetic data. The results for parameter retrieval were close to the ground truth, with the true values falling within one sigma of the posterior distribution. 
Regarding the data from the LBTI nuller, the model achieved accuracy comparable to the existing NSC methods. This indicates that the NPE model is highly effective for null depth retrieval, maintaining a level of precision at $10^{-4}$, which meets current standards in the field, with the benefit of a significantly faster inference.

An important strategy employed was fitting one model per pointing, rather than per star. This decision was driven by the fact that variations in observational conditions could affect the distribution of auxiliary data, thereby hindering the model's ability to generalize across different pointings. 
While this approach increased the time required for data generation and training, it also significantly enhanced the model's accuracy. Nonetheless, it remains faster than NSC in practice. The data simulation took approximately 60 seconds, while the model training required approximately 150 seconds (implemented on an NVIDIA A2 GPU and an AMD EPYC 7642 CPU). Since the model must regenerate data and retrain for each pointing, the data generation and training time must be multiplied by the number of pointings. For the six pointings in the dataset from the LBTI nuller, the total time for data generation and training is six times longer. Once trained, the model can be applied to all OBs within the given pointing.
During the inference process, the model completed the task in under two seconds, without the need for model reconstruction or data simulation. 
This represents a significant gain over classical techniques. Generating data and training for all OBs for a whole night takes approximately 30 minutes with NPE, compared to two hours with NSC\footnote{Private communication.}. 

Focusing on individual pointings, the model gained more specific experience with key data, resulting in improved performance and reliability. This step was crucial to ensure the model effectively handled the variability and complexity of observational data. These findings highlight that targeted training approaches can significantly enhance model performance.

\subsection{Trade-off between scalability and accuracy}
\label{sec:trade-off}

The results achieved by the current model successfully met our objective of testing NPE's ability for the reduction of nulling data. However, the optimal potential of this approach was not fully realised. Our ultimate goal was to develop a universal model capable of directly processing observational data and quickly retrieving null depth, without setting a different neural network per pointing.
The underlying model aims to provide a complete simulation environment for the physical processes. It samples parameters to be retrieved, as well as auxiliary data, from their priors to generate training data. Hence, it does not require the injection of real auxiliary data. This approach allows for a more flexible and generalisable model that can explore a wider range of scenarios.

\begin{figure}[ht]
    \centering
            \includegraphics[width=\linewidth]{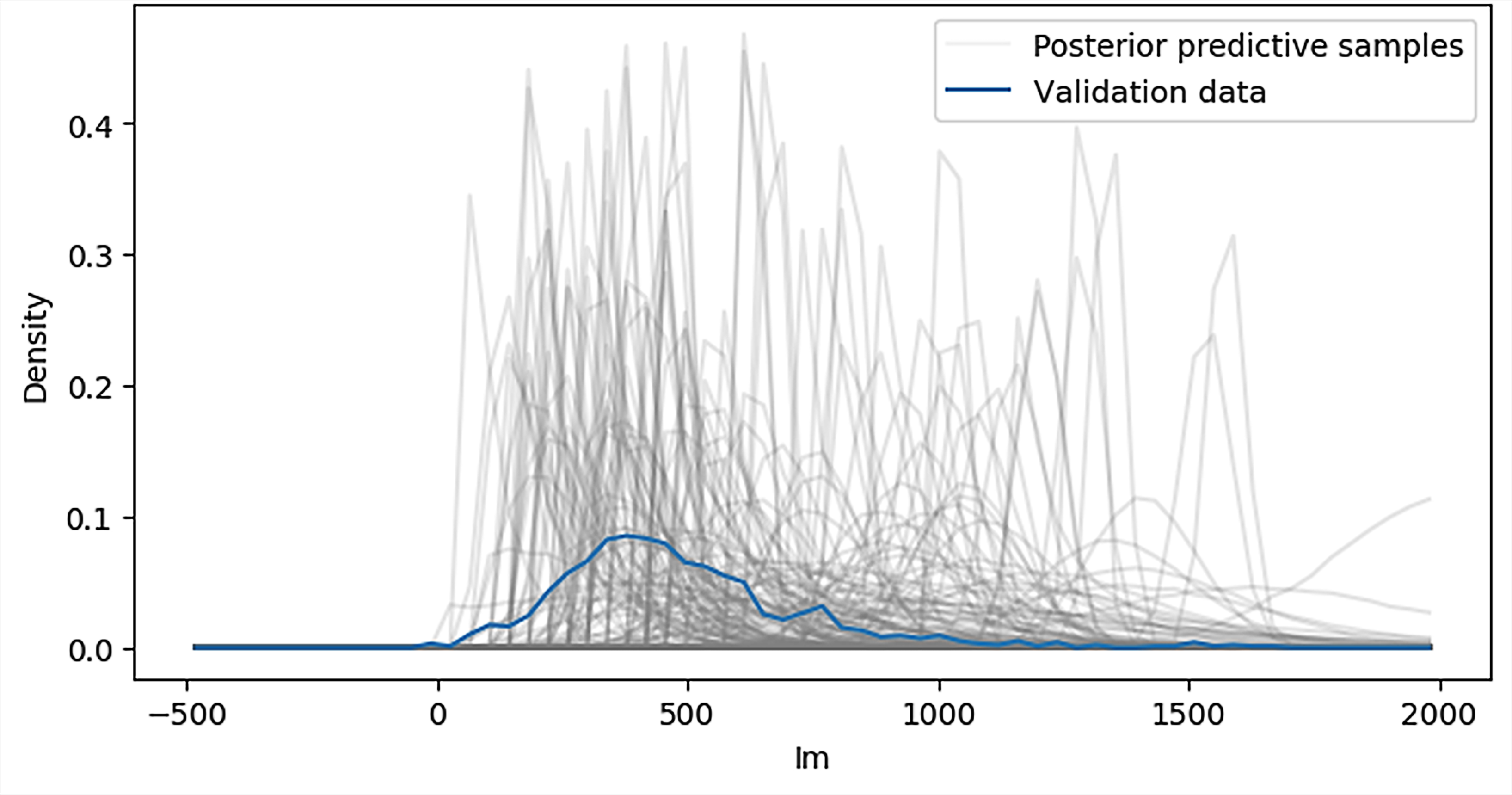}
    \caption{Posterior predictive check for the universal model. The grey semi-transparent lines represent the posterior predictive samples, while the solid blue line indicates the validation data.}
    \label{fig:ppc_comp}
    \end{figure}

In our experiments, we built and tested the “universal” model. Given priors for both the target parameters and the auxiliary data, this model generates simulations for all the measurements, and therefore produces training data that cover any situations potentially occurring in real observations, rather than being limited to a single pointing.
However, this universal model demonstrated inferior performance compared to the current pointing-specific model, with lower accuracy and greater uncertainty, both in the retrieval of synthetic data and LBTI data. 
The higher uncertainty is evident in the posterior predictive check for the universal model, as shown in Figure \ref{fig:ppc_comp}. We observed significant variability in the posterior predictive samples. The validation data fall within this wide range, suggesting that the model can generate data that spans the full spectrum of possible observations. This broad distribution could be attributed to the universal model's training on a diverse set of simulated data, which resulted in fewer similar data points during training and, consequently, a less focused posterior distribution. 

Consequently, we identified a trade-off between scalability and accuracy.
In our study, the current NPE model achieved high accuracy by focusing on individual pointings rather than a generalized model across all pointings.
Further studies are required to explore methods to overcome this trade-off, such as improving the data generation process to better reflect real-world observational conditions. Additionally, addressing challenges such as incorporating temporal correlations more effectively could enhance the model's ability to generalize across pointings without sacrificing accuracy. These potential improvements are discussed in the following section.

\subsection{Future improvements}
\label{sec:Limitations}

Several limitations were identified during the development and testing of our models, which need to be addressed to realise the potential of NPE for nulling interferometry fully. These limitations include the handling of temporal correlations, designing a better simulator or data preprocessing approach, and mitigating the threat of the simulation-to-reality gap.

\paragraph{Time correlations} 
A limitation of our present implementation is the reliance on marginal (histogram-based) distributions, which discard temporal and cross-channel correlations. In real-world observations, temporal correlations are unavoidable because consecutive measurements are inherently correlated in time during the data collection process \citep{norris2022optimal}. 
For instance, environmental changes and instrument stability over time can introduce temporal dependencies, which our current model does not fully capture. 
Therefore, a promising future direction is to improve the handling of time series data, where future models could experiment with embedding networks to manage the dimensionality of time series data \citep{vasist2023neural, dax2021real}.
This approach allows the model to learn temporal correlations and structures present in the data, potentially leading to more accurate retrievals. 
Embedding networks can transform high-dimensional time series data into a lower-dimensional space that preserves essential features, making it easier for subsequent models to process the data without losing crucial temporal information.

\paragraph{A better simulator} 
To enhance the model's performance, improving the simulator could provide a more accurate representation of the data. A possible approach is to use the aforementioned universal model and train a single neural network to infer all the OBs for the night.
One of the reasons for the failure of this model is the vastness of the parameter space, which makes it difficult to sample effectively.
In this model, the photometries of the beams are treated as independent from one another. 
However, this assumption leads to unrealistic simulations, such as one beam being very bright and the other very dim, which is not useful for training.
The diversity in the statistics of the residuals from the fringe tracker within an integration time, $\sigma_\epsilon$, is also challenging to simulate.
One potential solution is to use a hierarchical model, where the distributions of the intensities from the two beams are governed by a common "hyper-distribution," the parameters of which would need to be inferred. 
Such a distribution would actually model the fluxes emitted by the observed stars.
A similar hierarchy could be used to reflect the diversity of distributions of $\sigma_\epsilon$.

\paragraph{Simulation-to-reality gap}
A potential problem to be noted is that the current simulator relies heavily on the precision of the astrophysical model \citep{cannon2022investigating}. If the model does not accurately represent the actual data generation process, it can lead to significant issues in the analysis, creating a simulation-to-reality gap \citep{miglino1995evolving}. This could happen, for instance, in the presence of unknown systematic errors which are not captured by the theoretical model.
The current model is limited by the modelling of the thermal background \citep{defrere2016nulling}.
This limitation can be addressed through improved modeling or better preprocessing techniques to remove thermal background biases from the null signal dataset. 
A promising new method has been proposed to enhance the removal of thermal background biases \citep{Rousseau2024}.

Robust neural posterior estimation (RNPE, \citealt{ward2022robust}) offers a promising solution to this problem by incorporating an error model that accounts for discrepancies between simulated and observed data. This approach, referred to as "denoising," probabilistically inverts the error process of the observed data, enabling the detection and interpretation of discrepancies without confounding inference failures with simulator limitations. RNPE explicitly models the discrepancy between the observed data $y$ and the simulated data $x$ using an error model $p(y\vert x)$. 
This approach ensures that parameter estimates remain reliable even when the simulator is misspecified. By facilitating both model criticism and robust inference, RNPE provides a comprehensive framework for addressing the limitations of traditional NPE, making it well-suited for applications in null depth calibration where astrophysical models play a crucial role. 

\section{Conclusions}
\label{sec: conclusions}

In this paper, we explored the application of NPE for null depth retrieval in nulling interferometry. We focused on data obtained with the LBTI nuller and developed the model based on its underlying physical processes. The model was tested on both synthetic and observed data.

The results of the experiment demonstrated that the NPE model could infer parameters with high accuracy, providing precise estimates down to a few $10^{-4}$. This made it highly effective for null depth retrieval, achieving accuracy comparable to existing methods in practice. 
While the NPE approach provided results consistent with the classical self-calibration approach, it improved the error bar of one of the two scientific null measurements by a factor of three.
Furthermore, this approach offered improved computational efficiency via amortised inference.

This study also highlighted the trade-off between scalability and accuracy. The NPE model's approach of fitting one model per pointing rather than per star significantly enhanced the model's performance and reliability. 

Despite promising results, several limitations were identified, including challenges in handling temporal correlations, the need for a better simulator or data preprocessing methods, and the gap between simulation and reality. 
Future work should focus on addressing these limitations, to realise the full potential of NPE as well as applying this method to other nulling data sets obtained with the LBTI or other nullers.

Overall, this research demonstrates the potential of NPE for null depth retrieval in nulling interferometry, offering a scalable and accurate solution for astrophysical data reduction. This is particularly relevant in the framework of future nulling interferometers with more advanced beam combination schemes such as Asgard/NOTT \citep{defrere2024band} and LIFE \citep{2022A&A...664A..21Q}.
The advancements made in this study pave the way for further exploration and refinement of machine learning techniques in astrophysical applications, contributing to the broader field of exoplanet studies and beyond.

\begin{acknowledgements}
 D.D. and M.A.M have received funding from the European Research Council (ERC) under the European Union's Horizon 2020 research and innovation programme (grant agreement CoG - 866070). M-A.M. has received funding from the European Union’s Horizon 2020 research and innovation programme under grant agreement No.\ 101004719.
 The authors thank Prof. Olivier Absil and Dr. Malavika Vasist for their discussions that have started this project.
\end{acknowledgements}

\bibliographystyle{aa}
\bibliography{references}

\end{document}